\documentclass[12pt]{article}
\pdfoutput=1
\usepackage{amsmath}    	    
\usepackage[english]{babel}     
\usepackage{amssymb}
\usepackage{graphicx}   	    
\usepackage{float}
\usepackage{subfigure}
\usepackage{array}
\usepackage{color}      	    
\usepackage{hyperref}   	    
\usepackage[a4paper]{geometry}
\usepackage{caption}
\usepackage{listings}
\usepackage{comment}
\usepackage{tabularx}
\usepackage{graphicx}
\usepackage{adjustbox}
\usepackage{soul}

\newcolumntype{C}{>{\centering\arraybackslash}p{1.25cm}}

\def\lesim{\ \hbox to 0 pt{\raise .6ex\hbox{$<$}} \lower .5ex\hbox{$\sim$}\ }
\def\gesim{\ \hbox to 0 pt{\raise .6ex\hbox{$>$}} \lower .5ex\hbox{$\sim$}\ }
\def\insim{\ \hbox to 0 pt{\raise .5ex\hbox{$\in$}} \lower .6ex\hbox{$\sim$}\ }

\newcommand\unit[2]{\ifmmode \hbox{$#1$}\,\,{\rm #2} \else$\hbox{$#1$}\,\,{\rm #2}$\fi}
\newcommand\unitexp[3]{\ifmmode\hbox{$#1$}{\times 10^{#2}}\,\,{\rm #3}\else
$\hbox{$#1$}\times 10^{#2}\,\,{\rm #3}$\fi}

\newcommand{\pder}[2]{\frac{\partial#1}{\partial#2}}

\begin{document}

\title{LSTM-driven Forecast of CO$_{2}$ Injection in Porous Media}

\author{Gerald Kelechi Ekechukwu$^{1,3}$, Romain de Loubens$^{2}$ and Mauricio Araya-Polo$^{1}$\\
$^{1}$ TotalEnergies EP Research and Technology USA, Houston, Texas.\\
$^{2}$ TotalEnergies, 91120 Palaiseau, France.\\
$^{3}$ Louisiana State University, Baton Rouge, Louisiana.\\
mauricio.araya@totalenergies.com}

\date{\today}

\maketitle

\tableofcontents

\newpage

\begin{abstract}
   The ability to simulate the partial differential equations (PDE's) that govern multi-phase flow in porous media is essential for different applications such as geologic sequestration of CO$_2$, groundwater flow monitoring and hydrocarbon recovery from geologic formations \cite{Ekechukwu2021}. These multi-phase flow problems can be simulated by solving the governing PDE's numerically, using various discretization schemes such as finite elements, finite volumes, spectral methods, etc. More recently, the application of Machine Learning (ML) to approximate the solutions to PDE's has been a very active research area. However, most researchers have focused on the performance of their models within the time-space domain in which the models were trained. In this work, we apply ML techniques to approximate PDE solutions and focus on the forecasting problem outside of the training domain. To this end, we use two different ML architectures - the feed forward neural (FFN) network and the long short-term memory (LSTM)-based neural network, to predict the PDE solutions in future times based on the knowledge of the solutions in the past. The results of our methodology are presented on two example PDE's - namely a form of PDE that models the underground injection of CO$_2$ and its  hyperbolic limit which is a common benchmark case. In both cases, the LSTM architecture shows a huge potential to predict the solution behavior at future times based on prior data.
\end{abstract}

\section{Introduction}

The ability to simulate the governing partial differential equations (PDEs) that govern two-phase flow in porous media is essential for different applications such as geologic sequestration of CO$_2$ \cite{Ekechukwu2021}. Geological carbon sequestration which involves the injection of CO$_2$ into subsurface geologic formations is currently and actively being developed worldwide as a greenhouse mitigation strategy. Laboratory experiments, simulation studies, pilot projects suggest that large volumes of CO$_2$ can be safely stored and immobilized in the geologic formations for a long time.  Deep saline aquifers have been identified as one of such geologic formations with the greatest potential to store a substantial amount of this emitted greenhouse gases \cite{MICHAEL2010659}. Time-dependent PDEs have been used to model such flow through porous media.

Solving numerically non-linear and/or high dimensional PDE is frequently a challenging task and in most cases cannot be solved analytically, thereby depending heavily on numerical schemes. The development of accurate and efficient numerical schemes for solving such PDEs have been an active research area over the last few decades. \cite{courant1967,Cockburn2000,Thomas1995,johnson2012numerical}.More recently, the use of machine learning and deep learning to solve PDEs has emerged and is currently an active and vibrant area of research. Many researchers have utilized different approaches to try to solve the limiting hyperbolic case (Buckley-Leverett problem) of the time-dependent flow PDE using physics-informed neural network (PINN) \cite{fraces2020physics,Fuks_2020,gasmi2021physics}. The recent PINN studies show that the solution of the hyperbolic case which is solving a Riemann problem (i.e solution with discontinuity) is non-trivial. In this study, we used a purely data-driven approach motivated by \cite{hu2020neural}, to solve both the hyperbolic (Buckley-Leverett) and general(real) case studies. 
The novelties in this study include showing:
\begin{itemize}
    \item the use of LSTM for solving a Riemann problem. The time-dependent PDEs in \cite{hu2020neural} have no discontinuities.
    \item an investigation of the possibility of using smaller percentages of training data (between $40 \%$ and $80 \%$ ) to predict the future data.
    \item the use of simpler LSTM architecture to obtain very accurate results.
    \item the use of multivariate LSTM with no dependency on previous times or positions to predict future saturations. 
\end{itemize}

The main application is related to underground geological storage of CO$_2$, where preliminary coreflood experiments are required 
to characterize the two-phase flow properties of the formation where CO$_2$ is to be injected. The interpretation of such experiments 
yields parameter estimates that feed into large-scale models, aimed at predicting the CO$_2$ plume migration and its associated  
probability maps. 
Here a novel approach based on Deep neural networks (DNN) is proposed as an efficient and simple strategy to infer future CO$_2$ saturation profiles.

\section{Formulation of the CO$_{2}$ injection problem}

\subsection{Assumptions and notations}

\begin{figure}[H]
	\centering
	\includegraphics[width=0.6\linewidth]{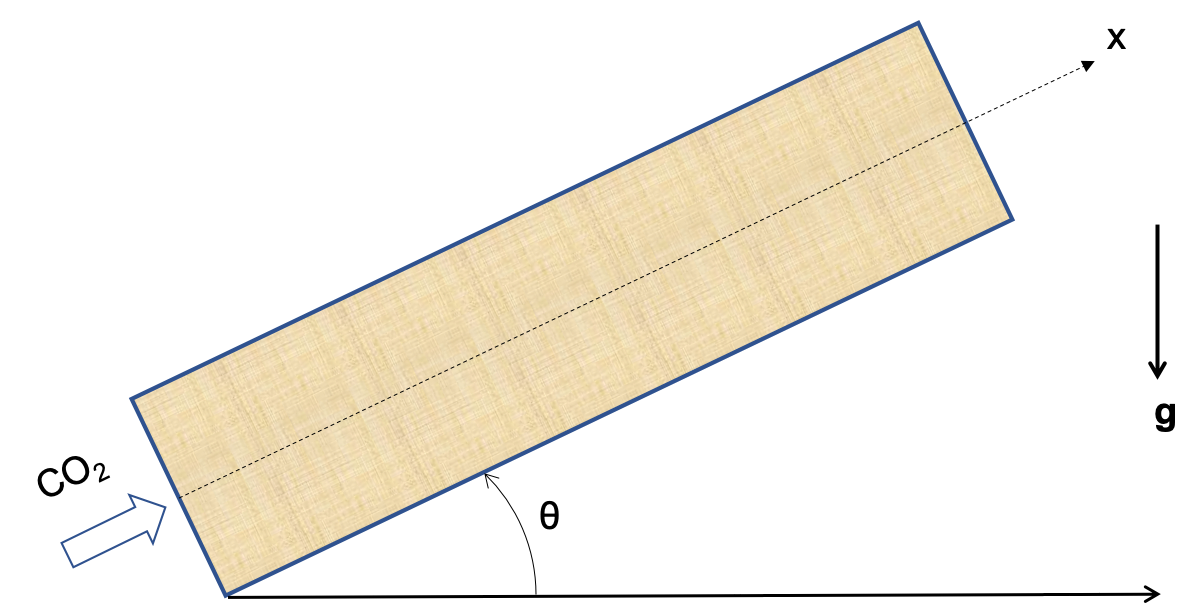}
	\caption{Schematic of a CO$_2$ coreflood experiment}
	\label{Fig:Coreflood}
\end{figure}

We consider the injection of CO$_2$ into a porous rock sample initially saturated with brine (i.e., saline water). The length of the rock sample, its cross-sectional area, porosity and permeability are denoted by $L$, $A$, $\phi$ and $k$, respectively.
The angle with respect to the horizontal direction is denoted by $\theta$; by convention, $\theta>0$ in the counter-clockwise direction (cf. Figure~\ref{Fig:Coreflood}).

The injection is modeled assuming a two-phase immiscible and incompressible displacement in 1D, under isothermal conditions. 
Typically, for the CO$_2$ sequestration application, the coreflood experiment would be carried out with supercritical CO$_2$, 
i.e., at $p>\unit{73.8}{bar}$ and $T>\unit{31}{^{\circ}C}$, and brine would be equilibrated with CO$_2$ before any injection takes place
\cite{krevor2012,Pini2013,bachu2013}.
In the following, subscripts $g$ and $w$ are used to denote the CO$_2$-rich 
and aqueous phases, respectively. Furthermore, we introduce the following notations and definitions:
\begin{itemize}
\item $p_j$ and $u_j$, the pressure and superficial velocity of phase $j$;
\item $\rho_j$, $\mu_j$ and $k_{rj}$, the density, viscosity and relative permeability of phase $j$;
\item $\sigma_{gw}$, the CO$_2$-water surface tension;
\item $p_c(S_g)=p_g - p_w$, the CO$_2$-water capillary pressure ($p_c'\ge 0$);
\item $M = \mu_w / \mu_g$, the viscosity ratio;
\item $G = k(\rho_w - \rho_g) g \sin(\theta) / (\mu_g u_T)$, the ratio of gravity to viscous effects;
\item $Ca = (\mu_g u_T L) / (\sigma_{gw} \sqrt{\phi k})$, the ratio of viscous to capillary effects (or capillary number);
\item $\lambda_j = k k_{rj}/ \mu_j$, the mobility of phase $j$;
\item $\lambda_T = \lambda_w + \lambda_g$, the total fluid mobility;
\item $f_j = \lambda_j / \lambda_T$, the fractional flow of phase $j$;
\item $F_g = \left(1+\frac{G}{M} k_{rw}\right)f_g$, the gravity-corrected CO$_2$ fractional flow;
\item $F_w = \left(1-G k_{rg}\right)f_w$, the gravity-corrected water fractional flow.
\end{itemize}

For phase $j=g,w$, the multiphase extension of Darcy’s law reads
\begin{equation}
\label{Eq:BL1D-Darcy-extended}
u_j = \lambda_j \left(-\pder{p_j}{x} - \rho_j g\sin(\theta)\right) \ , 
\end{equation}
and as a result of our 1D incompressible flow assumption, the total velocity $u_T=u_g+u_w$ is uniformly constant. Then we can derive the following velocity expressions,
\begin{eqnarray}
\label{Eq:BL1D-ug}
u_g &=& F_g u_T - \frac{\lambda_g \lambda_w}{\lambda_T}  \pder{p_c}{x} \ , \\
\label{Eq:BL1D-uw}
u_w &=& F_w u_T + \frac{\lambda_g \lambda_w}{\lambda_T}  \pder{p_c}{x} \ ,
\end{eqnarray}
where $F_g+F_w = f_g+f_w=1$.

\subsection{Unsteady-state experiment}

\subsubsection{Dimensional formulation}

Here we consider the unsteady flow experiment, when a constant CO$_2$ injection rate is imposed on the inlet face ($x=0$), 
so that $u_T = u_g = q_g/A$.

In addition to the velocity expressions of Eqs.~(\ref{Eq:BL1D-Darcy-extended}) to~(\ref{Eq:BL1D-uw}), we now have the continuity equation,
\begin{equation}
\label{Eq:BL1D-continuity}
0 = \phi \pder{S_j}{t} + \pder{u_j}{x}  \ ,
\end{equation}
which is a simple consequence of mass conservation within each phase $j=g,w$. We can then derive the following governing equations 
for water pressure, $p_w$, and CO$_2$ saturation, $S_g$: 
\begin{eqnarray}
\label{Eq:BL1D-pressure}
0 &=& \lambda_T \pder{p_w}{x} + u_T + (\lambda_g \rho_g + \lambda_w \rho_w)g\sin(\theta) + \lambda_g p_c' \pder{S_g}{x} \ , \\
\label{Eq:BL1D-saturation}
0 &=& \phi \pder{S_g}{t} + u_T \pder{F_g}{x} - \pder{}{x}\left(f_g \lambda_w p_c' \pder{S_g}{x}\right) \ .
\end{eqnarray}

At $t=0$, the CO$_2$ saturation in the core sample is zero and the water phase is in hydrostatic equilibrium 
(in particular $\lambda_g=u_T=0$ in Eq.~(\ref{Eq:BL1D-pressure})), hence we have the following initial condition,
\begin{equation}
\label{Eq:BL1D-IC}
S_g(x,0) = 0 \ , \quad p_w(x,0) = \bar{p} - \rho_w g \sin(\theta) (x - L) \ .
\end{equation}
where $\bar{p}$ is a reference pressure on the outlet face ($x=L$). 

From the condition $u_g=u_T$ on the inlet face, we have
\begin{equation}
\label{Eq:BL1D-BC-inlet}
-f_g \lambda_w p_c' \pder{S_g}{x} = (1-F_g) u_T \ , 
\end{equation}
which holds at $x=0$ and any time $t>0$. If $p_c' = 0$ then Eq.~(\ref{Eq:BL1D-BC-inlet}) implies $F_g=1$ (thus $S_g=1-S_{wr}$ if $G=0$), otherwise it yields
\begin{equation}
\label{Eq:BL1D-BC-inlet-bis}
-\pder{S_g}{x} = \left(\frac{1-G k_{rg}}{\lambda_g p_c'}\right) u_T\ .
\end{equation}
Note that $S_g=1-S_{wr}$ is always a solution of Eq.~(\ref{Eq:BL1D-BC-inlet}), as it yields 
$k_{rw}=\lambda_w=0$ and $F_g=f_g=1$; however, in the presence of  capillary effects this saturation may never be reached or is only reached after a long injection time. Therefore, unless capillary effects are neglected, 
Eq.~(\ref{Eq:BL1D-BC-inlet-bis}) should be used at $x=0$ instead of $S_g=1-S_{wr}$. 

On the outlet face ($x=L$), for each phase $j=g,w$ we impose
\begin{equation}
\label{Eq:BL1D-BC-outlet}
-\pder{p_j}{x}(L,t) =  \frac{p_j(L,t) - \bar{p}}{\delta} \ , 
\end{equation}
where $\delta$ is a fictitious layer thickness modeling the transition towards $p_c=0$ in the outflow vessel. Since our 
formulation is based on $p_w$ and $S_g$, the boundary condition for the CO$_2$ phase at $x=L$ is re-expressed as
\begin{equation}
\label{Eq:BL1D-BC-outlet-bis}
 - p_c'(S_g(L,t)) \pder{S_g}{x}(L,t) =  \frac{p_c(S_g(L,t))}{\delta} \ .
\end{equation}
Here $\delta$ should be considered as an uncertain parameter associated with the outflow condition, which can be adjusted to control the steady-state value of CO$_2$ saturation at the outlet.

In summary, the unsteady CO$_2$ coreflood experiment may be modeled by the following initial-boundary value problem (IBVP):
\begin{align}
\label{Eq:BL1D-IBVP-1}
& 0 = \lambda_T \pder{p_w}{x} + u_T + (\lambda_g \rho_g + \lambda_w \rho_w)g\sin(\theta) + \lambda_g p_c' \pder{S_g}{x} \ , \\
\label{Eq:BL1D-IBVP-2}
& 0 = \phi \pder{S_g}{t} + u_T \pder{F_g}{x} - \pder{}{x}\left(f_g \lambda_w p_c' \pder{S_g}{x}\right) \ , \\
\label{Eq:BL1D-IBVP-3}
& \text{(IC)} \quad S_g(x,0) = 0 \ , \quad p_w(x,0) = \bar{p} - \rho_w g \sin(\theta) (x - L)  \ , \\
\label{Eq:BL1D-IBVP-4}
& \text{(BC1)} \quad -p_c'(S_g(0,t))\pder{S_g}{x}(0,t) = \alpha(S_g(0,t))\, u_T \ , \\
\label{Eq:BL1D-IBVP-5}
& \text{(BC2)} \quad -p_c'(S_g(L,t)) \pder{S_g}{x}(L,t) =  \frac{p_c(S_g(L,t))}{\delta} \ , \\
\label{Eq:BL1D-IBVP-6}
& \text{(BC3)} \quad -\pder{p_w}{x}(L,t) =  \frac{p_w(L,t) - \bar{p}}{\delta} \ ,
\end{align}
where
\begin{equation}
\label{Eq:BL1D-alpha}
\alpha(S_g) = \frac{1-G k_{rg}(S_g)}{\lambda_g(S_g)} \ .
\end{equation}
Since the two unknowns $p_w$ and $S_g$ are only partially coupled, the governing PDE for $S_g$ (Eq.~(\ref{Eq:BL1D-IBVP-2})) may be solved first, with its initial condition $S_g(x,0) = 0$, and its nonlinear boundary conditions (BC1) and (BC2). Then knowing $S_g$, $p_w$ is easily computed from Eq.~(\ref{Eq:BL1D-IBVP-1}) and (BC3).

\subsubsection{Dimensionless formulation}

Here we introducing the classical Leverett $J$-function scaling, 
\begin{equation}
p_c(S_g) = \sigma_{gw} \sqrt{\frac{\phi}{k}} J(S_g) \ , 
\end{equation}
along with the following dimensionless variables:
\begin{itemize}
\item $x^* = x/ L$ ,
\item $t^* = u_T t / (\phi L)$, 
\item $\lambda_T^* = \mu_w\lambda_T/k = M k_{rg} + k_{rw}$, 
\item $p_w^* = (p_w + \rho_w g\sin(\theta)(x-L))k/(\mu_w u_T L)$.
\end{itemize}
For the unsteady flow problem, the non-dimensional form of Eqs.~(\ref{Eq:BL1D-IBVP-1}) to (\ref{Eq:BL1D-IBVP-6}) is
\begin{align} 
\label{Eq:BL1D-IBVP-nodim-1}
& 0 = \lambda_T^* \pder{p_w^*}{x^*} + 1 - G k_{rg} + \frac{1}{Ca}k_{rg}J' \pder{S_g}{x^*}\ , \\
\label{Eq:BL1D-IBVP-nodim-2}
& 0 = \pder{S_g}{t^*} + \pder{F_g}{x^*} - \frac{1}{MCa}\pder{}{x^*}\left(f_g k_{rw} J' \pder{S_g}{x^*}\right) \\
\label{Eq:BL1D-IBVP-nodim-3}
& \text{(IC)} \quad S_g(x^*,0) = 0 \ , \quad p_w^*(x^*,0) = \bar{p}^* \ , \\
\label{Eq:BL1D-IBVP-nodim-4}
& \text{(BC1)} \quad 0 = \frac{1}{Ca} J'(S_g(0,t^*))\pder{S_g}{x^*}(0,t^*) + \alpha^*(S_g(0,t^*)) \ , \\
\label{Eq:BL1D-IBVP-nodim-5}
& \text{(BC2)} \quad 0 = J'(S_g(1,t^*)) \pder{S_g}{x^*}(1,t^*) +  \frac{J(S_g(1,t^*))}{\delta^*} \ , \\
\label{Eq:BL1D-IBVP-nodim-6}
& \text{(BC3)} \quad 0 = \pder{p_w^*}{x^*}(1,t^*) + \frac{p_w^*(1,t^*) - \bar{p}_{\delta}^*}{\delta^*} \ ,
\end{align} 
where $\bar{p}^*=\bar{p}k/(\mu_w u_T L)$, $\bar{p}_{\delta}^*=(\bar{p}+\rho_w g\sin(\theta)\delta)k/(\mu_w u_T L)$ and
\begin{equation}
\label{Eq:BL1D-alpha-nodim}
\alpha^*(S_g) = \frac{1-G k_{rg}(S_g)}{k_{rg}(S_g)} \ .
\end{equation}
Using the above definition and Eq.~(\ref{Eq:BL1D-IBVP-nodim-1}), we verify that (BC1) is equivalent to $\pder{p_w^*}{x^*}(0,t^*)=0$,
which is consistent with the inlet condition $u_w=0$ (or $u_g=u_T$).

\subsubsection{Brooks-Corey expressions}

\begin{figure}[H]
	\centering
	\includegraphics[width=0.48\linewidth]{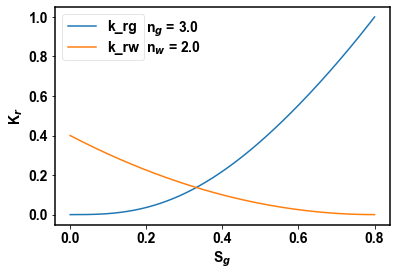}
	\includegraphics[width=0.48\linewidth]{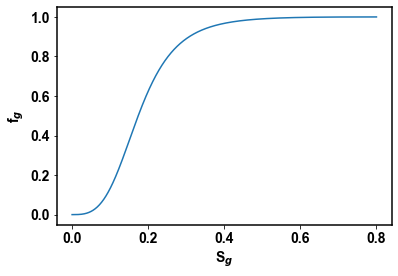}
	\includegraphics[width=0.5\linewidth]{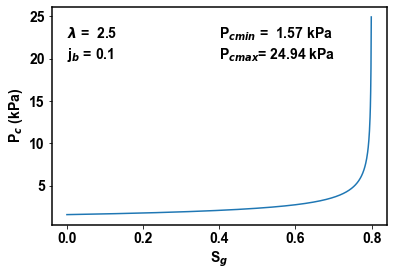}
	\caption{Corey-type relative permeabilities (top-left), flux function (top-right) and Brooks-Corey capillary pressure curves (bottom)}
	\label{Fig:krpc_Corey}
\end{figure}

Usually fluid parameters, such as densities and viscosities are well-known, whereas rock-fluid parameters, e.g., those used 
to represent relative permeability and capillary pressure curves, are uncertain. In particular, it is common for the type of 
coreflood experiment described above to use the following parametrizations, referred to as Brooks-Corey relative permeabilities 
and capillary pressure (cf. Fig.~\ref{Fig:krpc_Corey}),
\begin{eqnarray}
\label{Eq:BL1D-krg}
k_{rg} &=& k_{rg}^0 \left(1 - S_w^*\right)^2 \left[ 1 - (S_w^*)^{n_g}\right] \ , \\
\label{Eq:BL1D-krw}
k_{rw}  &=& k_{rw}^0 (S_w^*)^{n_w} \ ,  \\
\label{Eq:BL1D-pc}
p_c  &=&  j_b \,\sigma_{gw} \sqrt{\frac{\phi}{k}} \left(S_w^* \right)^{-1/\lambda}\ ,
\end{eqnarray}
where 
\begin{equation} 
\label{Eq:Sw-star}
S_w^* = \frac{S_w-S_{wr}}{1-S_{wr}}  \ .
\end{equation} 
Here $S_{wr}$ is the residual water saturation, $n_j$ and $k_{rj}^0$ are the Corey exponent and the maximum relative permeability 
of phase $j$, $\lambda$ is a pore-size distribution index, and $j_b$ is a dimensionless scaling factor for the CO$_2$ drainage entry 
pressure.

The Brooks-Corey parameters that will be used in this study are provided in Table \ref{coreyparamtable}. Furthermore, the rock and fluid properties are shown in Table \ref{rftable}.

\begin{table}[H]
\centering
\begin{tabular}{||c c c ||} 
 \hline
 Parameters & Hyperbolic Case & General Case  \\ [0.8ex] 
 \hline\hline
 $k_{rg}^0$ & 1.0 & 1.0 \\ 
 \hline
 $k_{rw}^0$ & 1.0 & 0.4  \\
 \hline
  $n_{g}$ & 2.0 & 3.0  \\
 \hline
 $n_{w}$ & 2.0 & 2.0  \\
 \hline
 $S_{wr}$ & 0.0 & 0.2  \\
 \hline
 $j_{b}$ & -  & 0.1  \\
 \hline
 $\lambda$ & - & 2.5  \\
 \hline
 \end{tabular}
 \caption{Corey and Brooks-Corey parameters used in this study (for the general case)}
\label{coreyparamtable}
\end{table}

\begin{table}[H]
\centering
\begin{tabular}{||c c || c c ||} 
 \hline
Rock Properties and dimensions &  & Fluid Properties &   \\ [0.8ex] 
 \hline\hline
Length (m) & 0.1 & Surface tension (mN/m) & 32  \\ 
 \hline
Diameter (m) & 0.0508 & Gas viscosity $\mu_{g}$ (Pa.s) & 2.3e-5   \\ 
 \hline
Porosity (-) & 0.221 & Water viscosity $\mu_{w}$ (Pa.s) & 5.5e-4  \\ 
 \hline
 Permeability ($m^2$) & 0.912e-12 & Total flow rate ($m^3$/min) & 15  \\ 
 \hline
 \end{tabular}
 \caption{Rock and fluid properties}
\label{rftable}
\end{table}

\section{Machine Learning  Approach}
\label{sec:ml}
In this section, we describe the Machine Learning (ML) approach that will be applied to approximate the PDE solution for the above described problem. Before addressing the selected ML architecture, we present the construction of the dataset used for training, validation and testing.

\subsection{Dataset generation}

The datasets are computed using either the analytical solution (hyperbolic case) of the initial-boundary-value problem or in the general case, its numerical solution computed by a Finite Volume (FV) solver implemented by a proprietary implementation. In the following description, we will only consider the horizontal case ($\theta=0$, hence $G=0$). \\

The data generation process by numerical simulation on a fixed grid can be summarized as:
$$FV(\Theta, x_i, t^n) \to \left\{u_i^n\right\}$$
where $\Theta$ represents the parameters listed in the previous section, $x_i$ and $t^n$ represent location and time in the discrete space-time domain. The output of $FV(\Theta,x_i,t^n)$ is the numerical approximation of gas saturation at grid position $i$ and timestep $n$, denoted as $u_i^n$.

The FV solver implements a time-explicit scheme, with upstream weighting for the convective term and edge-centered approximation of the diffusive term. Considering a uniform grid of size $m$, the resulting discretization of Eq.~\ref{Eq:BL1D-IBVP-nodim-2}, for $1\le i\le m$, is the following :
\begin{equation}
u_i^{n+1} = u_i^n + \frac{\Delta t}{\Delta x}\left(f_g(u_{i-1}^n)-f_g(u_i^n)\right) + \frac{\Delta t}{M Ca \Delta x}\left(\mathcal{F}_{i+1/2}^n-\mathcal{F}_{i-1/2}^n\right)\ , 
\end{equation}
where 
\begin{eqnarray}
\mathcal{F}_{i+1/2}^n =
\left\{ \begin{array}{cc} 
-\frac{Ca}{k_{rg}(u_0^n)} f_{g}(u_0^n) k_{rw}(u_0^n) \ , & i=0 \ , \\ \\
\frac{1}{4}\left(f_g(u_{i+1}^n) + f_g(u_i^n)\right) \left(k_{rw}(u_{i+1}^n) + k_{rw}(u_i^n)\right) \frac{J(u_{i+1}^n) - J(u_i^n)}{\Delta x} \ , & 1\le i\le m-1 \ , \\ \\
-\frac{1}{\delta^*} f_{g}(u_{m+1}^n) k_{rw}(u_{m+1}^n) J(u_{m+1}^n) \ , & i=m \ . 
\end{array} \right. 
\end{eqnarray}
Arrays $\mathbf{u}^n$ (resp. $\mathbf{u}^{n+1}$) are of size $m+2$ and contain the discrete values of gas saturation at timestep $n$ (resp. $n+1$). Index $i=0$ (resp. $i=m+1$) corresponds to a ghost cell where the inlet (resp. outlet) saturation value $u_0^n$ (resp. $u_{m+1}^n$) is stored and updated according to the boundary conditions. More precisely, we obtain $u_0^{n+1}$ (resp. $u_{m+1}^{n+1}$) from $u_1^{n+1}$ (resp. $u_m^{n+1}$) using the discrete form of Eqs.~(\ref{Eq:BL1D-IBVP-nodim-4}) (resp. Eq.~(\ref{Eq:BL1D-IBVP-nodim-5})) and an iterative procedure (e.g., dichotomy followed by Newton's method) to invert the resulting nonlinear relationships.

\begin{figure}[H]
	\centering
    \includegraphics[scale = 0.7]{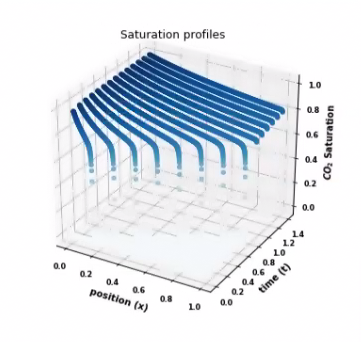}
    \includegraphics[scale = 0.48]{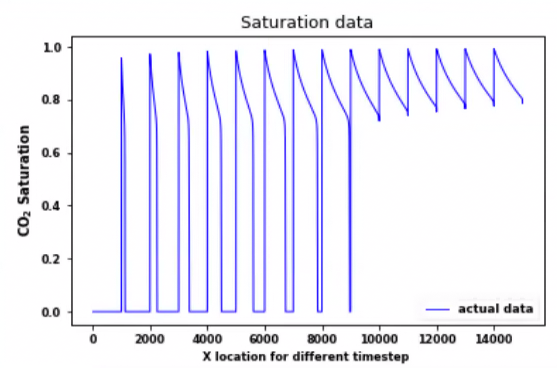}
    \caption{Saturation data in 3D and 2D projections}
    \label{Fig:Sat_data}
\end{figure}

The spatio-temporal coordinates are mapped to the saturation values as obtained from the numerical scheme using different train/test splits during the training. For the results presented in following sections, we use datasets of size $15000$ samples in both the hyperbolic case and the general case, where each sample is composed of ($x_i$, $t^n$) as input $X$, and ($u_i^n$) as output $Y$ (or label). The split between training and testing data will be specified along each numerical experiment. \textcolor{black}{Figure \ref{Fig:Sat_data} (left) shows the 3D view of the saturation data in which each profile belongs to a specific time. Each saturation profile contains 1000 saturation values corresponding to 1000 discrete spatial points. The 2D projection is a compressed view in which saturation profiles at consecutive timesteps are stacked together.} 

\begin{figure}[H]
    \includegraphics[scale = 0.31]{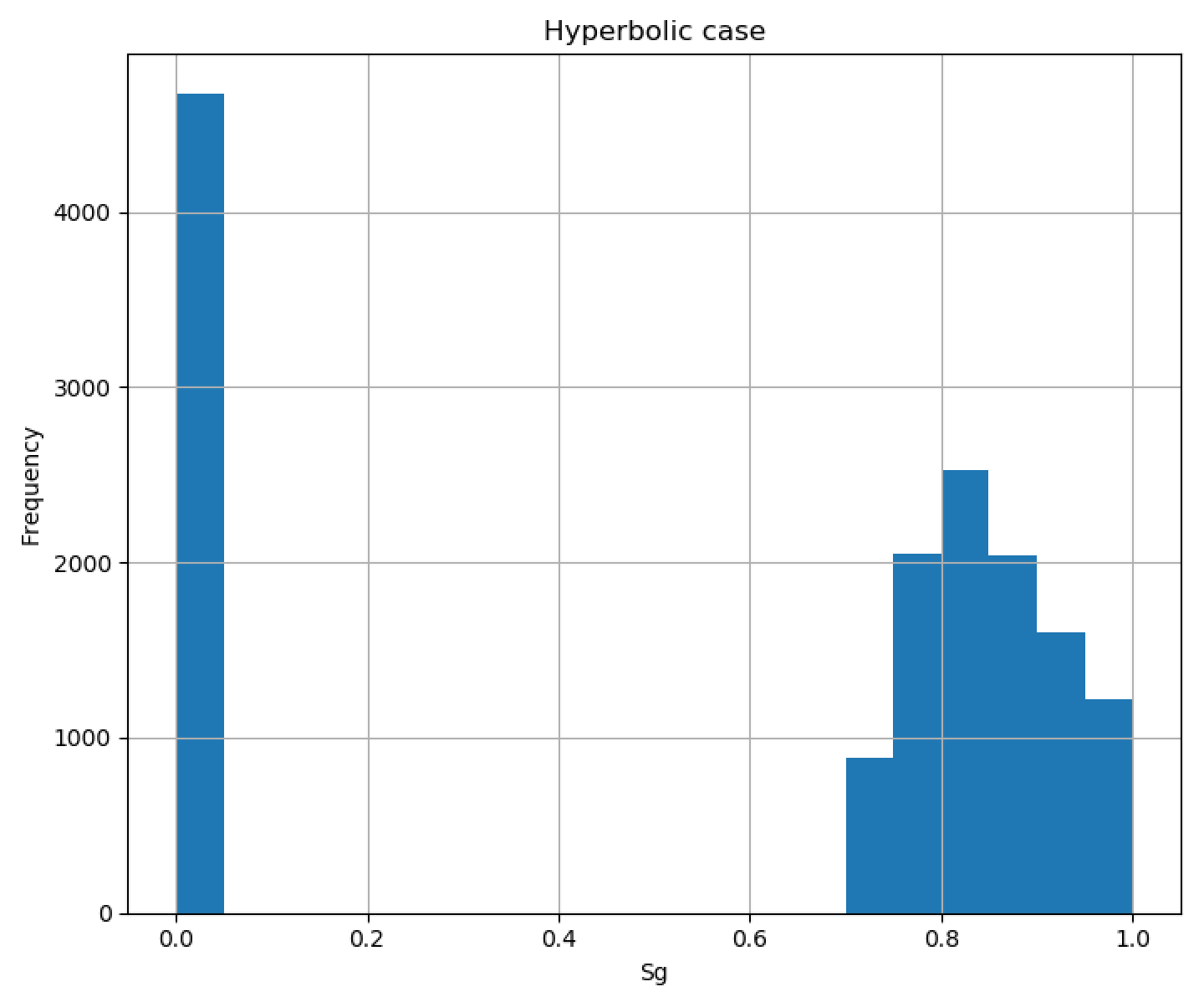}
    \includegraphics[scale = 0.31]{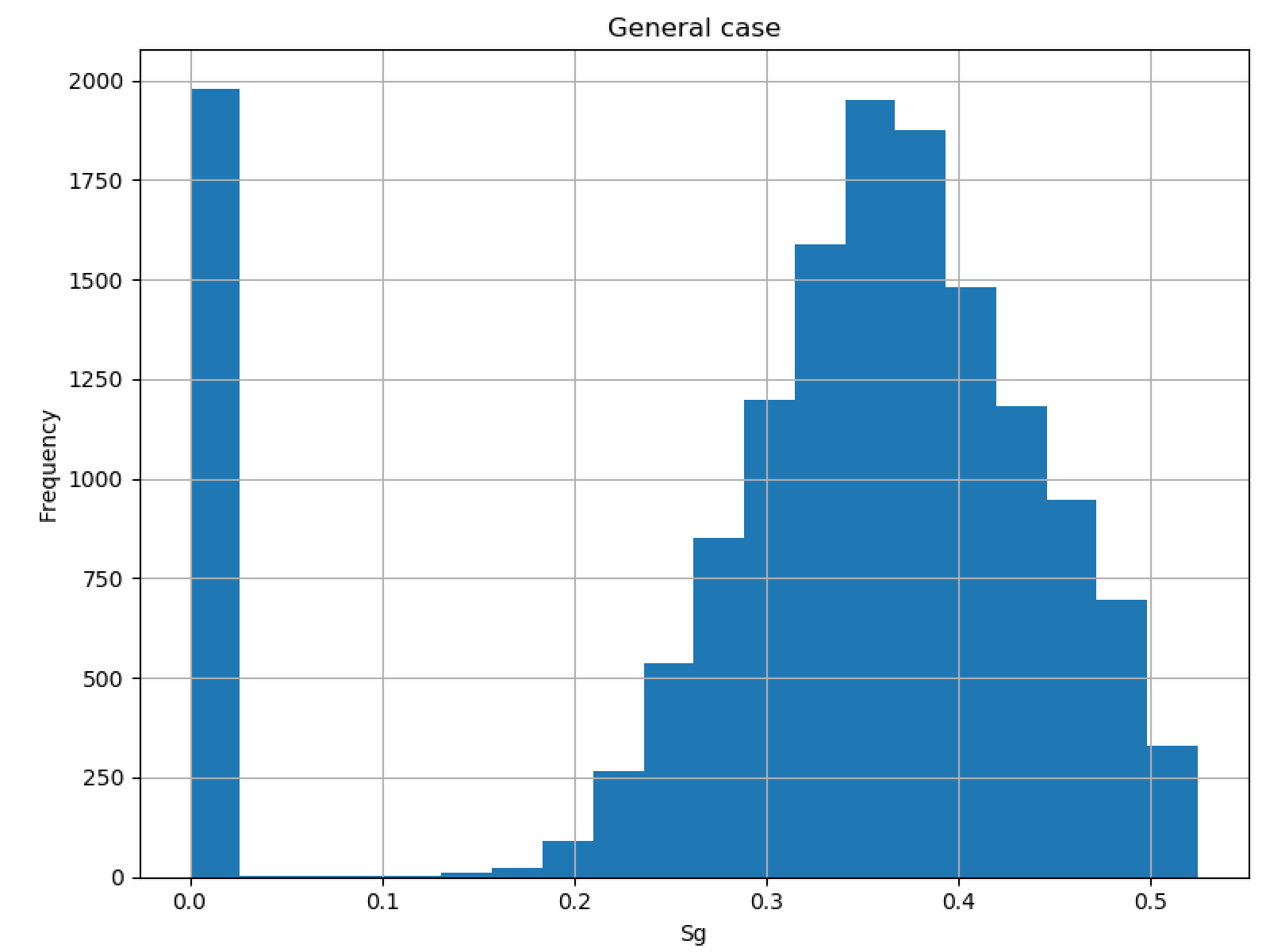}
    \caption{Distribution of computed $S_{g}$ for both hyperbolic and general case.}
    \label{Fig:data_dist}
\end{figure}


\subsection{Machine Learning architectures}

Two main classes of ML architectures were deployed during the research, traditional Feed Forward Networks (FFN) and Long-short Term Memory (LSTM) based Networks.

\subsubsection{Feed Forward Network architectures}
Artificial neural networks are a biologically-inspired computational models which consists of a set of artificial neurons (called nodes or units) and a set of directed edges between them. This computational neurons are represented by non-linear activation functions, weights and biases. In a supervised setting, learning consists on minimizing the loss between prediction and ground-truth, the weights are iteratively modified by back-propagation of errors from the loss function, this till the training converges to a minimum loss.\\ 
Typically, neurons are depicted as circles or squares and the edges as the arrows connecting them. Each edge is usually associated with a weight $w_{jj'}$ from node $j'$ to node $j$ using the "to-from" notation. Associated with each neuron $j$ is an activation function $l_j$ $(.)$. Hence, the value $v_j$ of each neuron $j$ is computed by applying a non-liner activation function on the weighted sum of its inputs $v_j = l_j(\sum_{j'} w_{jj'} . v_j')$. Feed forward networks (FFN) are networks that avoid cycles in a directed graph of nodes. In FFN, all nodes are typically arranged into layers and the outputs in each layer can be calculated using the outputs from the previous layers as shown below. In this work, we use a class of FNNs called Multi-layer Perceptron (MLP).

\begin{figure}[H]
	\centering
    \includegraphics[scale = 0.9]{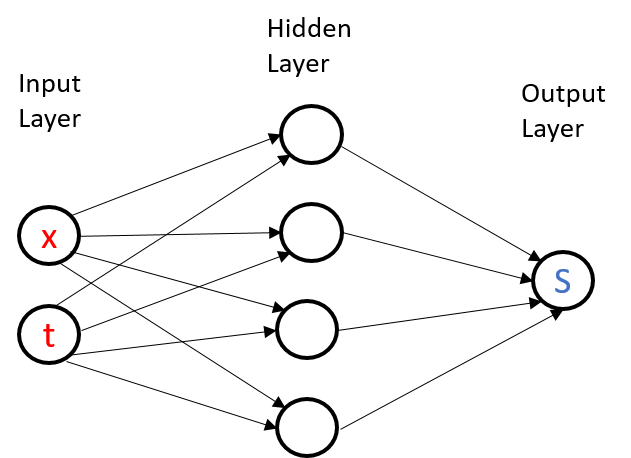}
    \caption{Sample of feed forward network architecture}
    \label{Fig:fnn_arch}
\end{figure}

In our study, the spatio-temporal coordinates are mapped to the saturation values as obtain from the numerical scheme using different train/test splits during the FFN model training. First, we tested different number of hidden layers and neurons in which hidden layer. Different hidden layers $[5,8,10,15,20]$ and neuron-per-layer $[5,10,15,20,25,30,40,50]$ combinations were tried, it was found the best performing hidden layer/layer-neuron combination to be 5/50 for this problem. Both manual and automatic search algorithms were employed in an attempt to discover the optimum hyperparameters for each model. In ML, there are several choices of the activation function but we found the \textit{tanh} activation function to be the best for this Buckley-Leverett (BL) problem. The \textit{Adam} optimization algorithm was used to train (i.e., minimize the error in) the network. After the network training is completed, it can be used to forecast the saturation values for the rest of the timesteps.


\subsubsection{LSTM-based Network architectures}


Recurrent neural networks (RNN) are a special class of artificial neural networks that captures dependencies in the data by providing a feedback loop in the network. They are memory networks in that the output of the current state partially depends on the the outputs of the previous states as well. 
RNNs have been widely used for natural language processing, audio and speech processing \cite{Chien} as well as video processing \cite{Ma18} or even seismic inversion \cite{adler18}.
RNNs are supposed to also theoretically learn long-term dependencies but they fail due to vanishing or exploding gradients especially for long sequences. New RNN architectures have been designed to overcome this problem and one of such architecture is the Long Short-Term Memory (LSTM). We would be using LSTM is this study, hence we would briefly describe the inner workings next.

\begin{figure}[H]
	\centering
    \includegraphics[scale = 0.5]{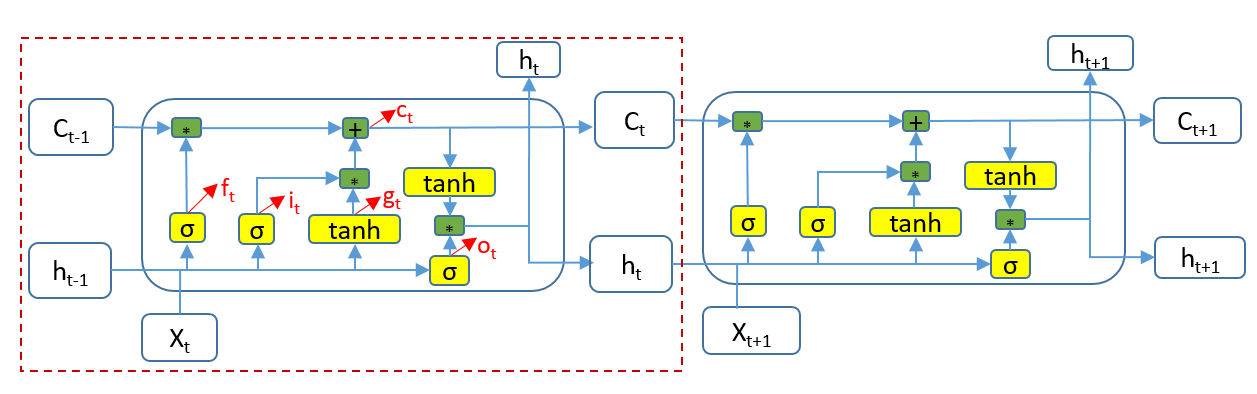}
    \caption{LSTM Architecture description}
    \label{Fig:lstm_arch}
\end{figure}

The RNN involves the mapping from an input sequence ($x_1,x_2,..., x_T $) to an output target sequence ($y_1,y_2,..., y_T $) where each data point $x_t$ or $y_t$ is a real-valued saturation vector.  A training set is usually a set of samples where each sample is an (input sequence, output sequence) pair where either the input or output could also be a single data point \cite{Lipton15}.
At each time step, the LSTM cell has three inputs - the short term memory from the previous timestep or the hidden state ($h_{t-1}$), the long term memory ($c_{t-1}$) and the current input data ($X_{t}$) - and two outputs - the short-term memory ($h_{t}$) from the current time step and the new cell state ($C_{t}$).
To further understand the inner workings of the LSTM method, intermediate computations are described briefly as follows \cite{SAK18} and the network unit activation are calculated iteratively from $t=1$ to $T$.

\begin{equation} 
\label{Eq:f-t}
f_{t} = \sigma (W_{if}x_t + b_{if} + W_{hf}h_{t-1} + b_{hf} ) 
\end{equation} 

\begin{equation} 
\label{Eq:i-t}
i_{t} = \sigma (W_{ii}x_t + b_{ii} + W_{hi}h_{t-1} + b_{hi} ) 
\end{equation} 

\begin{equation} 
\label{Eq:g-t}
g_{t} = tanh (W_{ig}x_t + b_{ig} + W_{hg}h_{t-1} + b_{hg} ) 
\end{equation}

\begin{equation} 
\label{Eq:o-t}
o_{t} = tanh (W_{io}x_t + b_{io} + W_{ho}h_{t-1} + b_{ho} ) 
\end{equation}

\begin{equation} 
\label{Eq:c-t}
c_{t} = f_t \odot c_{t-1} + i_t \odot g_{t}  
\end{equation}

\begin{equation} 
\label{Eq:h-t}
h_{t} = o_t \odot tanh (c_{t} )
\end{equation}

Where $f$, $i$, $o$, $c$  are the forget gate, input gate, output gate and cell state vectors respectively, $W$ is the weight matrices (e.g. $W_{if}$ is the weight matrix from the input to the forget gate) , $b$ connotes the bias terms, $\sigma$ and tanh are respectively the logistic $sigmoid$ function and $tanh$ activation function, $\odot$ is the element-wise product of the vectors.
The input gate decides what new information should be stored in the long-term memory and takes as input, the information from the current input and the short-term memory from the previous step. The forget gate decides which information from the long-term memory/cell state should be discarded. The output gate uses the newly computed long-term memory, the previous short-term memory/hidden state and the current input to produce the new hidden state which is then passed on to the LSTM cell in the next time step.

\begin{figure}[H]
	\centering
	\includegraphics[width=0.6\linewidth]{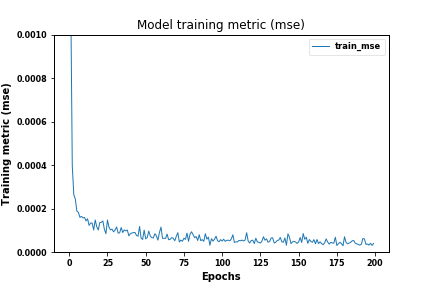}
    \caption{ Performance metric as training progresses. Last MSE value is $0.00012$ at epoch $200$.}

	\label{Fig:LSTM_loss}
\end{figure}

With respect to our application, in order to preserve the unique functional mapping from one set of ($x_i$,$t_i$), to ($s_i$), only one time-stepping pre-processing approach was used. This means that the saturation values are only dependent on its corresponding position and time values and not on any previous values of $x_i$,$t_i$, $s_i$. Two fully connected (FC) layers were added to 3-layer LSTM block selected for this study. The first FC (100-neuron) layer was added before the LSTM block while the second FC (1-neuron) layer was added after the LSTM block. The architectural design is the result of a parametrical search towards accuracy. \textcolor{black}{The LSTM layers with activation of "tanh" and recurrent activation of "sigmoid" trained to an epoch of 200 gave the best performing result. Figure \ref{Fig:LSTM_loss} for example, shows how the performance of the learning process as the training progresses. The optimum results for the general case using LSTM and MLP were obtained when the epoch was 200 but for the hyperbolic case using MLP the epochs had to be increased to 800. Above these mentioned epochs, the performances of the test data deteriorated.}

\section {Results and Discussion}

\subsection{Forecasting with Supervised Learning}

In this section, we present the results of the prediction/forecasting capabilities of our ML models. The saturation dataset obtained using the numerical scheme is split into training and test samples. The percentage of the training data ranges from $40\%$ to $80\%$ which corresponds to the early time steps or prior data of the time range considered in this ML problem. The rest of the dataset is then used for predicting performance of the trained models. Furthermore, the first $10\%$ of the training data contains the initial condition of the PDE solution.



We first trained the networks to develop a functional relationship between the input ($x_1,x_2,..., x_T $) and target output ($y_1,y_2,..., y_T $) sequences. The mean square error $(MSE)$ between the predicted output ($\hat{y_1},\hat{y_2},..., \hat{y_T} $) and the actual output sequences ($y_1,y_2,..., y_T $) were used as the loss function to update the weights and biases during the training.  


\textcolor{black}{The performance of the prediction saturations were compared using three metrics - the root mean square error (RMSE) between the actual and predicted profiles, the coefficient of correlation ($R^2$ score) between the actual and predicted profiles as well as  the image-based intersection over union of the pixels predicted compared to the actual. The $R^2$ and $RMSE$ were computed both profile-wise (i.e. for each predicted saturation profile) as well as overall for all predicted saturation profiles. The profile- wise performance metrics are shown in the tables while the overall test data metrics is shown on the figures. The $R^2$ scores usually range from - to 1. The closer to 1, the score is, the better the match between the predicted and actual saturation values. For the RMSE, a better perfomance is seen when the RMSE is comparatively smaller. The IoU is commonly defined as the ratio of true positives to the sum of true positives, false positives and false negatives. IoU also ranges from 0 to 1 with 1 signfying a perfect match and 0 signifying  no match.} In the following, we show results for both the hyperbolic and general cases.

\subsubsection{Hyperbolic case}
\begin{figure}[H]
	\centering

	\includegraphics[width=0.48\linewidth]{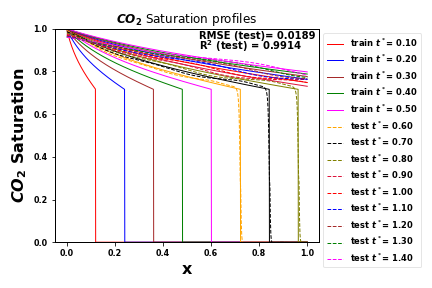}
	\includegraphics[width=0.48\linewidth]{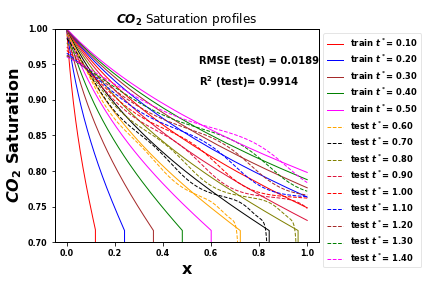}
	\includegraphics[width=0.48\linewidth]{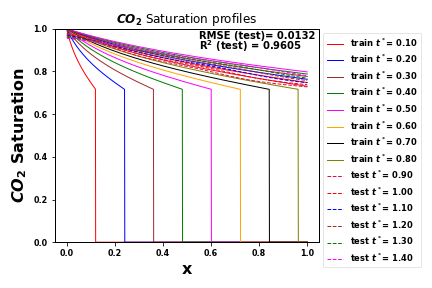}
	\includegraphics[width=0.48\linewidth]{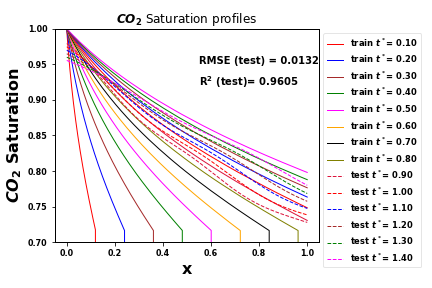}
    \includegraphics[width=0.48\linewidth]{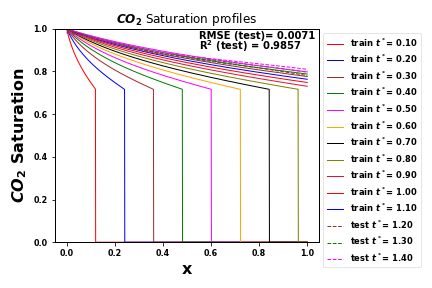}
	\includegraphics[width=0.48\linewidth]{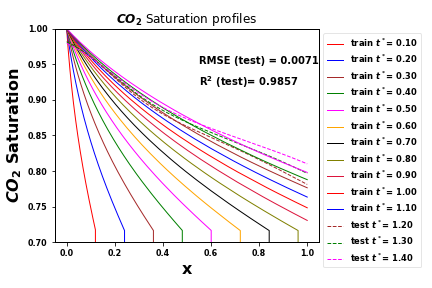}

    \caption{forward prediction in time using MLP/FFN: with $40\%$ prior data for training (top-left), with corresponding zoomed plot (top-right)  , with $60\%$ prior data for training (middle-left), with corresponding zoomed plot  (middle-right) , with $80\%$ prior data for training (bottom-left), with corresponding zoomed plot  (bottom-right)  }

	\label{Fig:M1_FFN}
\end{figure}

\paragraph {MLP results:}
Figure \ref{Fig:M1_FFN} shows the results for the forward prediction in time using the FFN. On the top row, we show the results when the first $40\%$ of the dataset was used for training. This implies that saturation profiles at times $t^*$ = [0,0.1,0.2,0.3,0.4,0.5] were used for training while the saturation profiles from $t^*$ = [0.6 to 1.4] were inferred.Amongst the inferred data, only the saturation profiles at times 0.6, 0.7 and 0.8 contained both the rarefaction and shock components. A good match was obtained for $t^*$ = 0.6 with an $R^2$ score of 0.995. It should be noted that in all figures presented, the inferred saturation profiles are shown in "dashed lines" while the actual profiles are shown in "solid lines". The $R^2$ between the actual and the predicted profiles decreased while the RMSE increased as the time progressed further away from the training times. Furthermore, it can be seen that the $R^2$ increases with more training data. For instance, the last saturation profile i.e. at $t^*$ =1.4 has an $R^2$ of 0.9305 when $40\%$ data was used compared to 0.979 when the training data was $80\%$. The full performance metrics for all the inferred saturation profiles at $40\%$, $60\%$ and $80\%$ are respectively  shown in Table \ref{40p_MLP},  Table \ref{60p_MLP} and  Table \ref{80p_MLP}. \textcolor{black}{Most of the mismatch were seen in the rarefaction parts of the saturation profiles. Also, the predicted inlet gas saturation at different times varied from 0.95 and 1 instead remaining fixed at $S_g = 1$.}

\begin{table}[]
\centering
\begin{adjustbox}{width=\textwidth}
\begin{tabular}{||c  |c c | c c ||} 
\hline
 & General case && Hyperbolic case & \\ [0.8ex] 
 \hline
 Saturation profile at & RMSE & $R^2$ & RMSE & $R^2$ \\ [0.8ex] 
  \hline\hline
 $t^* = 0.6$ & 0.00718 & 0.98422 & 0.02636 & 0.99514 \\ 
 \hline
 $t^* = 0.7$ & 0.00885 & 0.97542 & 0.02498 & 0.99357  \\
 \hline
  $t^* = 0.8$& 0.01099 & 0.96141 & 0.02673 & 0.97708  \\
 \hline
  $t^* = 0.9$ & 0.01347 & 0.94122 & 0.01198 & 0.97381 \\ 
 \hline
 $t^* = 1.0$ & 0.01617 & 0.91444 & 0.01261 & 0.96707  \\
 \hline
  $t^* = 1.1$& 0.01901 & 0.88092 & 0.01391 & 0.95506   \\
 \hline
 $t^* = 1.2$ & 0.02195 & 0.84056 & 0.01493 & 0.94238  \\
 \hline
 $t^* = 1.3$ & 0.02497 & 0.79326  & 0.01515 & 0.93451  \\
 \hline
 $t^* = 1.4$& 0.02805  & 0.73896 & 0.01489 & 0.93054  \\
  \hline \hline
 All test data $t^*$ s & Intersection over Union (IoU) && Intersection over Union (IoU) & \\ 
  \hline  \hline
 & 0.930 && 0.947 & \\ 
 \hline
 
 \end{tabular}
  \end{adjustbox}
\caption{Test data performance metrics with $40\%$ training data using MLP}
\label{40p_MLP}
\end{table}


\begin{table}[H]
\centering
\begin{adjustbox}{width=\textwidth}
\begin{tabular}{||c  |c c | c c ||} 
\hline
 & General case && Hyperbolic case & \\ [0.8ex] 
 \hline
 Saturation profile at & RMSE & $R^2$ & RMSE & $R^2$ \\ [0.8ex] 
  \hline\hline
 $t^* = 0.9$ & 0.00938 & 0.97148 & 0.00945 & 0.98370 \\ 
 \hline
 $t^* = 1.0$ & 0.01109 & 0.95971 & 0.01107 & 0.97462  \\
 \hline
  $t^* = 1.1$& 0.01311 & 0.94335 & 0.01247 & 0.96386   \\
 \hline
 $t^* = 1.2$ & 0.01529 & 0.92263 & 0.01367 & 0.95173  \\
 \hline
 $t^* = 1.3$ & 0.01723 & 0.89810  & 0.01491 & 0.93656  \\
 \hline
 $t^* = 1.4$& 0.01976  & 0.87038 & 0.01639 & 0.91588  \\
  \hline \hline
 All test data $t^*$ s & Intersection over Union (IoU) && Intersection over Union (IoU) & \\ 
  \hline  \hline
 & 0.947 && 0.963 & \\ 
 \hline
 
 \end{tabular}
  \end{adjustbox}
 \caption{Test data performance metrics with $60\%$ training data using MLP}
\label{60p_MLP}
\end{table}

\begin{table}[H]
\centering
\begin{adjustbox}{width=\textwidth}
\begin{tabular}{||c  |c c | c c ||} 
\hline
 & General case && Hyperbolic case & \\ [0.8ex] 
 \hline
 Saturation profile at & RMSE & $R^2$ & RMSE & $R^2$ \\ [0.8ex] 
  \hline\hline
 $t^* = 1.2$ & 0.00923 & 0.97179 & 0.00599 & 0.99072  \\
 \hline
 $t^* = 1.3$ & 0.01098 & 0.95999  & 0.00703 & 0.98592  \\
 \hline
 $t^* = 1.4$& 0.01294  & 0.94446 & 0.00817 & 0.97909  \\
  \hline \hline
 All test data $t^*$ s & Intersection over Union (IoU) && Intersection over Union (IoU) & \\ 
  \hline  \hline
 & 0.971 && 0.981 & \\ 
 \hline
 \end{tabular}
 \end{adjustbox}
 \caption{Test data performance metrics with $80\%$ training data using MLP}
\label{80p_MLP}
\end{table}

\paragraph{Hybrid-LSTM results:}

\begin{figure}[H]
	\centering

	\includegraphics[width=0.48\linewidth]{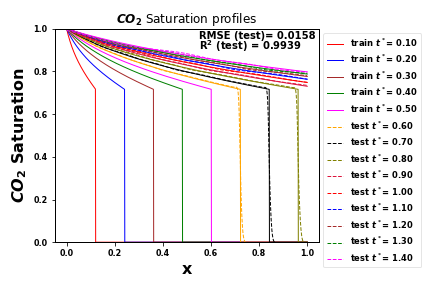}
	\includegraphics[width=0.48\linewidth]{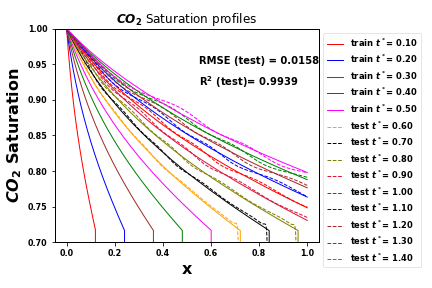}
	\includegraphics[width=0.48\linewidth]{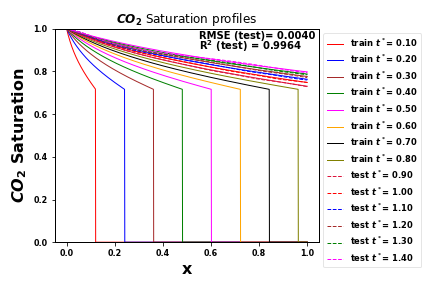}
	\includegraphics[width=0.48\linewidth]{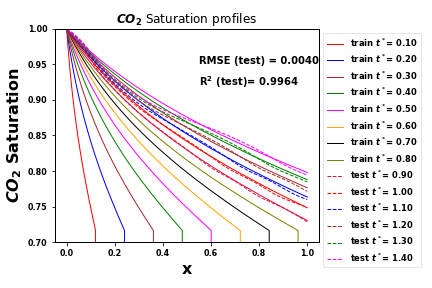}
    \includegraphics[width=0.48\linewidth]{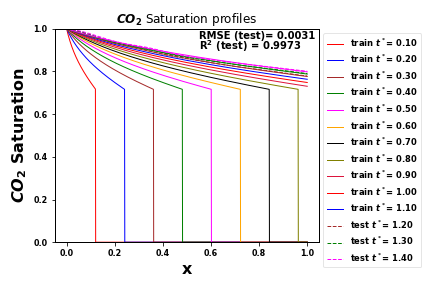}
	\includegraphics[width=0.48\linewidth]{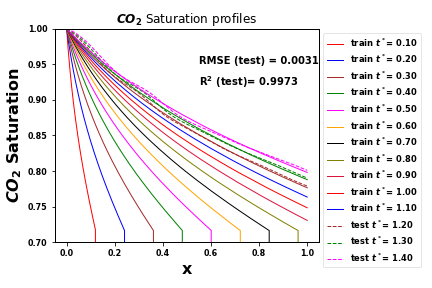}

    \caption{forward prediction in time using Hybrid-LSTM: with $40\%$ prior data for training (top-left), with corresponding zoomed plot (top-right)  , with $60\%$ prior data for training (middle-left), with corresponding zoomed plot  (middle-right) , with $80\%$ prior data for training (bottom-left), with corresponding zoomed plot  (bottom-right)  }

	\label{Fig:M1_LSTM}
\end{figure}

Figure \ref{Fig:M1_LSTM} shows the predicted saturation results using the Hybrid-LSTM model. The inferred results showed a better performance  than those obtained using the MLP models. The overall test performance metric at the different training data sample size in terms of $R^2$ score ranged from 0.9605 to 0.9914 when the MLP is used but when the Hybrid-LSTM is used, the $R^2$ score ranged from 0.9939 to 0.9973.
The full performance metrics for all the inferred saturation profiles at $40\%$, $60\%$ and $80\%$ are respectively  shown in Table \ref{40p_Hybrid-LSTM},  Table \ref{60p_Hybrid-LSTM} and  Table \ref{80p_Hybrid-LSTM}. Similar to the MLP results, we see that the saturation profiles at times closer to those of the training times are better than those further away.

\begin{table}[H]
\centering
\begin{adjustbox}{width=\textwidth}
\begin{tabular}{||c  |c c | c c ||} 
\hline
 & General case && Hyperbolic case & \\ [0.8ex] 
 \hline
 Saturation profile at & RMSE & $R^2$ & RMSE & $R^2$ \\ [0.8ex] 
  \hline\hline
 $t^* = 0.6$ & 0.00720 & 0.98415 & 0.02616 & 0.99522 \\ 
 \hline
 $t^* = 0.7$ & 0.00880 & 0.97567 & 0.02681 & 0.99259  \\
 \hline
  $t^* = 0.8$& 0.00998 & 0.96819 & 0.02768 & 0.97533  \\
 \hline
  $t^* = 0.9$ & 0.01101 & 0.96074 & 0.00280 & 0.99857 \\ 
 \hline
 $t^* = 1.0$ & 0.01210 & 0.95206 & 0.00283 & 0.99834  \\
 \hline
  $t^* = 1.1$& 0.01332 & 0.94157 & 0.00313 & 0.99772   \\
 \hline
 $t^* = 1.2$ & 0.01463 & 0.92919 & 0.00354 & 0.99676  \\
 \hline
 $t^* = 1.3$ & 0.01601 & 0.91495  & 0.00398 & 0.99547  \\
 \hline
 $t^* = 1.4$& 0.01744  & 0.89906 & 0.00465 & 0.99324  \\
  \hline \hline
 All test data $t^*$ s & Intersection over Union (IoU) && Intersection over Union (IoU) & \\ 
  \hline  \hline
 & 0.961 && 0.969 & \\ 
 \hline
 \end{tabular}
  \end{adjustbox}
 \caption{Test data performance metrics with for $40\%$ training data using Hybrid-LSTM}
\label{40p_Hybrid-LSTM}
\end{table}

\begin{table}[H]
\centering
\begin{adjustbox}{width=\textwidth}
\begin{tabular}{||c  |c c | c c ||} 
\hline
 & General case && Hyperbolic case & \\ [0.8ex] 
 \hline
 Saturation profile at & RMSE & $R^2$ & RMSE & $R^2$ \\ [0.8ex] 
  \hline\hline
 $t^* = 0.9$ & 0.00688 & 0.98465 & 0.00207 & 0.99922 \\ 
 \hline
 $t^* = 1.0$ & 0.00914 & 0.97268 & 0.00227 & 0.99894  \\
 \hline
  $t^* = 1.1$& 0.01101 & 0.96007 & 0.00296 & 0.99796   \\
 \hline
 $t^* = 1.2$ & 0.01244 & 0.94879 & 0.00380 & 0.99627  \\
 \hline
 $t^* = 1.3$ & 0.01365 & 0.93820  & 0.00492 & 0.99309  \\
 \hline
 $t^* = 1.4$& 0.01482  & 0.92710 & 0.00622 & 0.98787  \\
  \hline \hline
 All test data $t^*$ s & Intersection over Union (IoU) && Intersection over Union (IoU) & \\ 
  \hline  \hline
 & 0.957 && 0.984 & \\ 
 \hline
 \end{tabular}
  \end{adjustbox}
 \caption{Test data performance metrics with $60\%$ training data using Hybrid-LSTM}
\label{60p_Hybrid-LSTM}
\end{table}

\begin{table}[H]
\centering
\begin{adjustbox}{width=\textwidth}
\begin{tabular}{||c  |c c | c c ||} 
\hline
 & General case && Hyperbolic case & \\ [0.8ex] 
 \hline
 Saturation profile at & RMSE & $R^2$ & RMSE & $R^2$ \\ [0.8ex] 
  \hline\hline
 $t^* = 1.2$ & 0.00437 & 0.99369 & 0.00253 & 0.99835  \\
 \hline
 $t^* = 1.3$ & 0.00649 & 0.98601  & 0.00302 & 0.99739  \\
 \hline
 $t^* = 1.4$& 0.00905  & 0.97282 & 0.00363 & 0.99588  \\
  \hline \hline
 All test data $t^*$ s & Intersection over Union (IoU) && Intersection over Union (IoU) & \\ 
  \hline  \hline
 & 0.984 && 0.993 & \\ 
 \hline
 \end{tabular}
 \end{adjustbox}
 \caption{Test data performance metrics with $80\%$ training data using Hybrid-LSTM}
\label{80p_Hybrid-LSTM}
\end{table}

\subsubsection{General case}
\begin{figure}[H]
	\centering

	\includegraphics[width=0.48\linewidth]{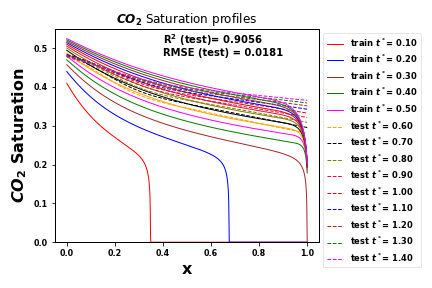}
	\includegraphics[width=0.48\linewidth]{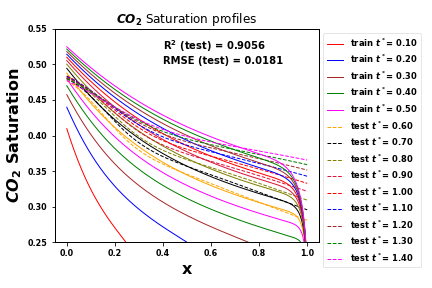}
	\includegraphics[width=0.48\linewidth]{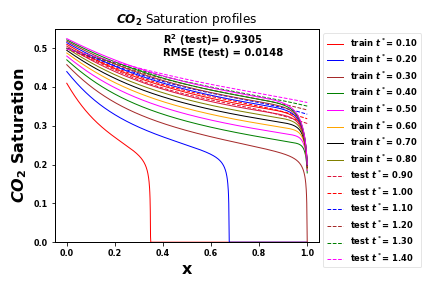}
	\includegraphics[width=0.48\linewidth]{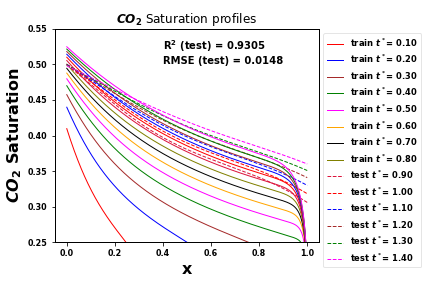}
    \includegraphics[width=0.48\linewidth]{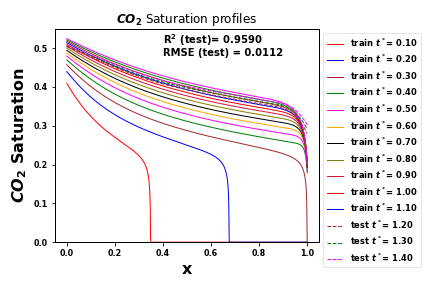}
	\includegraphics[width=0.48\linewidth]{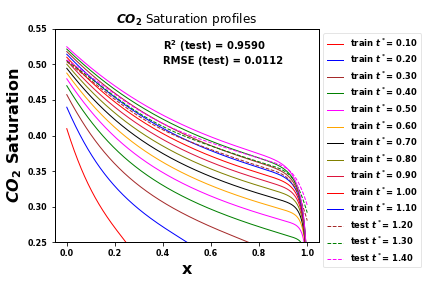}

    \caption{forward prediction in time using MLP/FFN: with $40\%$ prior data for training (top-left), with corresponding zoomed plot (top-right)  , with $60\%$ prior data for training (middle-left), with corresponding zoomed plot  (middle-right) , with $80\%$ prior data for training (bottom-left), with corresponding zoomed plot  (bottom-right)  }

	\label{Fig:M24_FFN}
\end{figure}

By general case we mean the case in which the capillary pressure is used during the PDE formulation. The main implication of using capillary pressure is that first, it is a more realistic scenario and closer to what is observed in actual physical experiments. The saturation profiles are impacted by diffusion and exhibit a smeared transition between the rarefaction and shock front. Additionally, the capillary end-effect toward the end of the core sample (at $x=1$) is clearly visible in the actual numerical data (shown in thick lines) in Figure \ref{Fig:M24_FFN}.

\paragraph {MLP results:}
Figure \ref{Fig:M24_FFN} shows the results for the forward prediction in time using the FFN for this more realistic case study. When the first $40\%$ of the data was used for training, the overall prediction performance in terms of $R^2$ was 0.9056. This performance metrics increased to 0.9305 and 0.9590 when the percentage training data increased to $60\%$ and $80\%$, respectively. On a saturation profile-wise analysis, the inferred profiles at the times closest to those of the training data gave high $R^2$ values than those that are further away. For instance, using $60\%$ training data, the $R^2$ at $t^*$ = 0.9 was 0.971 while that at $t^*$ = 1.4 was 0.870. From figure \ref{Fig:M24_FFN}, we see that the capillary-end portion parts were not captured when the percentage training data were either $40\%$ or $60\%$. However, when the training dataset was increased to $80\%$, there was improvement in the ability of the model to mimic the end effects

\begin{figure}[H]
	\centering

	\includegraphics[width=0.48\linewidth]{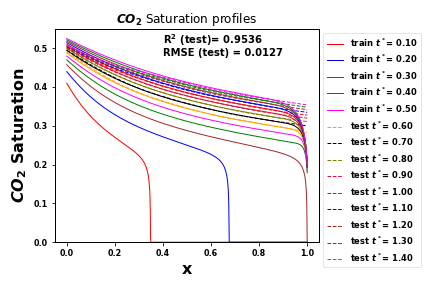}
	\includegraphics[width=0.48\linewidth]{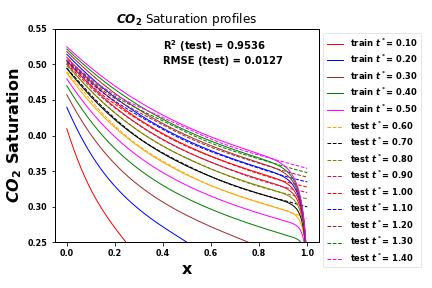}
	\includegraphics[width=0.48\linewidth]{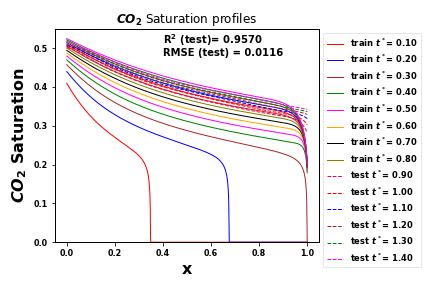}
	\includegraphics[width=0.48\linewidth]{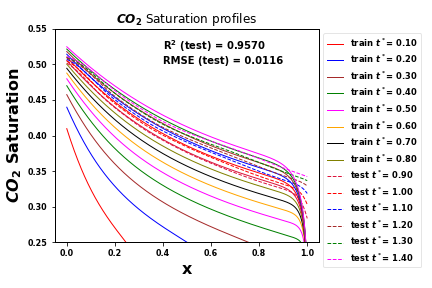}
    \includegraphics[width=0.48\linewidth]{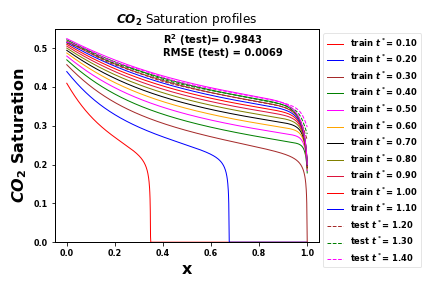}
	\includegraphics[width=0.48\linewidth]{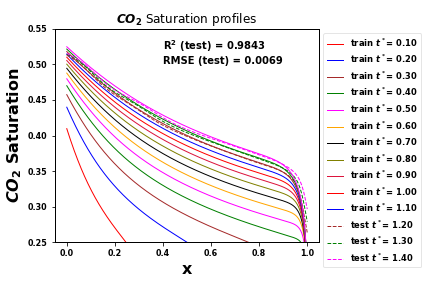}

    \caption{forward prediction in time using Hybrid-LSTM: with $40\%$ prior data for training (top-left), with corresponding zoomed plot (top-right)  , with $60\%$ prior data for training (middle-left), with corresponding zoomed plot  (middle-right) , with $80\%$ prior data for training (bottom-left), with corresponding zoomed plot  (bottom-right)  }
	\label{Fig:M24_LSTM}
\end{figure}

\paragraph{Hybrid-LSTM results:}
Figure \ref{Fig:M24_LSTM} shows the results for the forward prediction in time using the Hybrid-LSTM model. The inferred profiles showed higher $R^2$ scores than their corresponding profiles obtained from the MLP model. However, using $40\%$ and $60\%$ of the data for training was still not sufficient to capture the capillary end effects. The capillary effects were slightly captured when $60\%$ training data was used but fully captured when the training data was increased to $80\%$. \textcolor{black}{Since the LSTM-based model gave better results in terms of the quantitative performance metrics as well as a better visual match, we should use it for further extensions of the work}

\begin{figure}[H]
	\centering
	\includegraphics[width=0.9\linewidth]{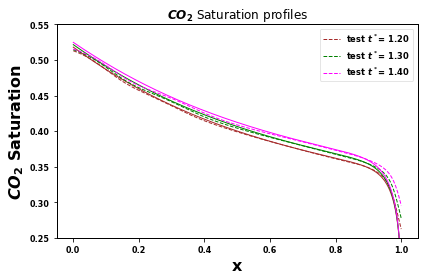}

		\caption{Forward prediction for $t^* = 1.2, 1.3, 1.4$ using Hybrid-LSTM (general case) - the training data ranges from $t^*=0$ to  $t^*=1.1$. Solid curves represent the ground-truth.}

	\label{Fig:M1_LSTM_last_add}
\end{figure}

Figures \ref{Fig:M1_LSTM_last_add} is a zoomed plot of figure \ref{Fig:M24_LSTM} (last row) above without the training data included in the same plot.As with other plots, the same colors corresponds to same timesteps while the dashed line is for the predicted profile, the solid line is that of the actual profile.

\begin{figure}[H]
	\hspace{-0.6in}
	\includegraphics[width=1.1\linewidth]{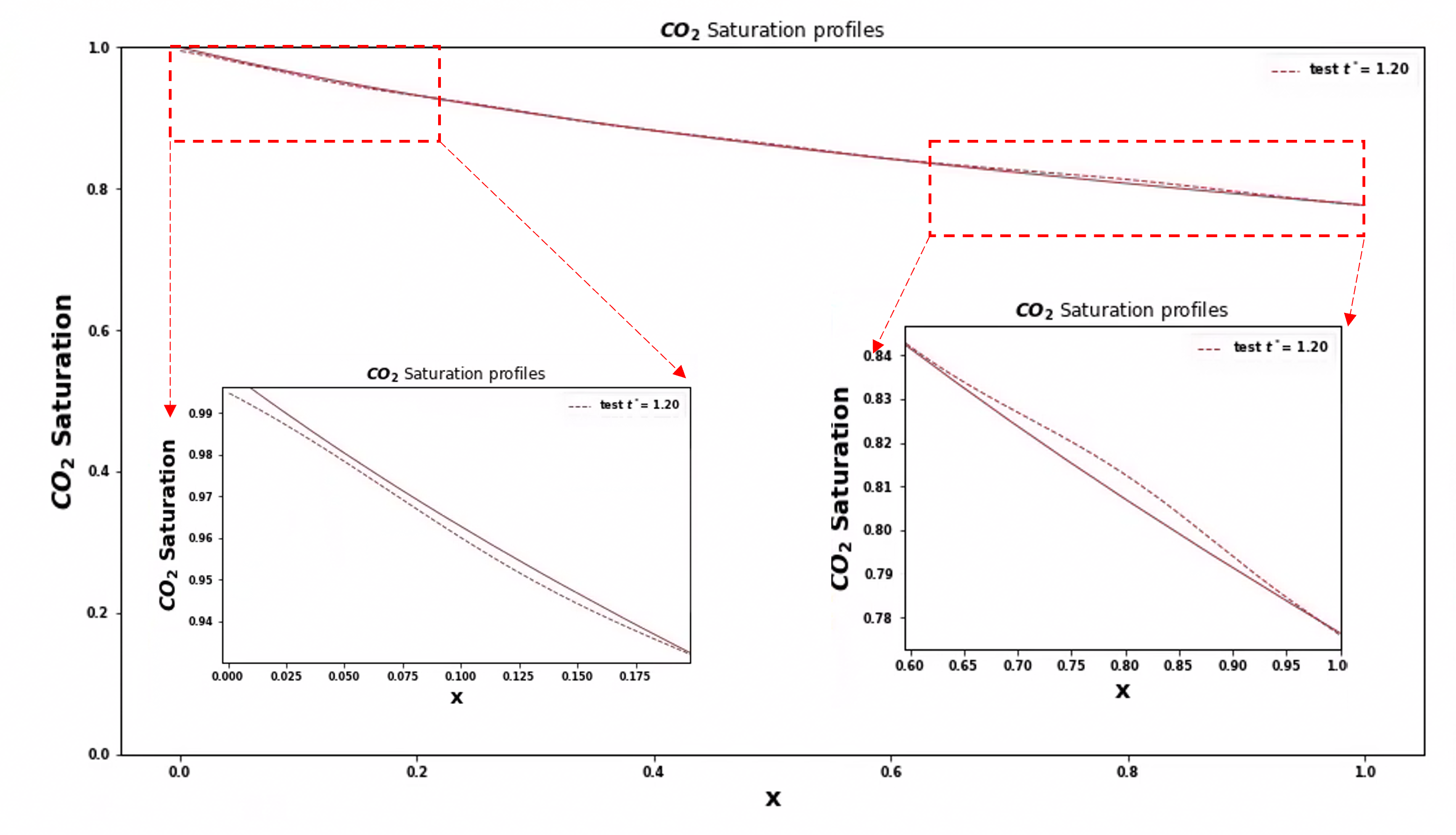}

		\caption{Forward prediction for $t^* = 1.2$ using Hybrid-LSTM (Hyperbolic case) - the training data ranges from $t^*=0$ to  $t^*=1.1$. The insets (in red dashed-line rectangles) show the zoom plots at the inlets and outlets of the core.}

	\label{Fig:M1_LSTM_single}
\end{figure}

\begin{figure}[H]
	\hspace{-0.8in}
	\includegraphics[width=1.1\linewidth]{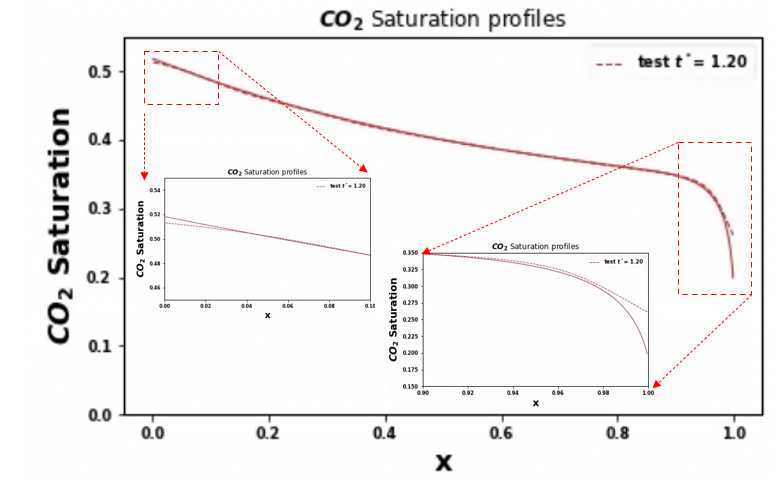}

		\caption{Forward prediction for $t^* = 1.2$ using Hybrid-LSTM (General case) - the training data ranges from $t^*=0$ to  $t^*=1.1$. In other words, with $80\%$ training data. The insets (in red dashed-line rectangles) show the zoom plots at the inlets and outlets of the core.}

	\label{Fig:M24_LSTM_single}
\end{figure}

Figures \ref{Fig:M1_LSTM_single} and \ref{Fig:M24_LSTM_single} each shows a profile of one of the predicted saturation profiles when $80\%$ of the training data was used for forward prediction for the hyperbolic and general case respectively. The insets in the figures displays a zoom of the inlet and outlet portions of the saturation profile. The difference in the profiles is less than 0.005 saturation units for the hyperbolic case.For the general case the difference is more pronounced at the outlet than the inlet, but even the maximum difference in predicted versus actual saturation value is about 0.075 saturation units.

\subsection {Alternative architectures and approaches}
As part of this internship, different machine and deep learning techniques were implemented. They can be summarized as listed below.
\begin{itemize}
\item Many different architectures of MLP were tested with hidden layers varying from $1$ to $30$ and nodes per layer varying from $5$ to $200$, also with different activation functions and optimization schemes. 
\item Many LSTM-based variations combining hidden layers, nodes per layers, activation functions were tested.
\item Different architectures of hybrid-LSTM with various combinations of LSTM hidden layers, dense hidden layers, nodes per layers, activation functions were tested.
\item Different architectures of Bi-directional LSTM.
\item Neural networks with 1D and 2D convolutional layers were also implemented with different filter sizes, number of filters and depths tried.
\item Neural networks starting with convolutional layers followed by LSTM layers  were also implemented with different filter sizes, number of filters and depths tried.
\item \textcolor{black}{Also implemented were tree-based machine learning techniques such as decision trees, random forest, gradient boosting etc.as well as other conventional ML techniques such as Support vector machines (SVM) and Least squares methods with different kernels were implemented. These conventional ML approaches were not able to capture the trends in the solution of the PDE.}
\item Finally, a Physics Informed Neural Network (PINN) approach failed to solved the general case, which include realistic capillary pressure. Description of this effort is out of the scope of this work.
\end{itemize}


None of these gave a better result than the Hybrid-LSTM model architecture presented.

\section {Conclusion}

In this work, we investigated the possibility of using machine learning for predicting saturation profiles at future timesteps. The datasets used for this work were generated using either the analytical solution (hyperbolic case) of the initial-boundary-value problem or in the general case, its numerical solution computed by a Finite Volume (FV) solver. In the machine learning framework, some part of the dataset (known as training data) is used for training and optimizing the model parameters while the rest (called test data) is used for testing the performance of the model. We varied the percentage of our training data between $40\%$ and $80\%$ such that the training set corresponds to the dataset in the prior time steps. The aim is to predict the saturation profiles in future timesteps. We emphasize that most of the previous purely data-driven approaches have focused on the interpolation or ‘‘training domain" problem. By this we mean that the saturations predicted were within the range of the $position$ and $time$ input values that was used for training. Our study is also different from the approach of \cite{hu2020neural} that used a univariate LSTM approach. They used previous saturation values to predict future saturation values. In the formulation of their test data, the input to the trained model were chosen from the actual data which may not necessarily be the correct approach. In our study, we used both $position$ and $time$ as inputs (which is similar to how the numerical schemes are formulated as well) and we are predicting a unique saturation value for each input values. Hence, there is no dependence on the previous saturation (output) or previous $position$ and $time$ values.

With our formulation, we present two deep neural network models (feed forward network and LSTM-based models) that are capable of predicting future saturation profiles. The LSTM-based network gave an overall better performance than the feed forward network. As a perspective for future work, we will investigate LSTM-based inversion of 1D core flood experimental data, i.e., inference of the rock and fluid properties that were considered as inputs to the simulated datasets. Future work may also include extensions to two and three spatial dimensions.\\

\section* {Acknowledgements}
We would like to thank TotalEnergies for permission to share this work. Also, the authors are grateful to Adrien Lemercier who implemented the FV solver for the general case (with capillary pressure) during a summer internship in 2020.

\bibliography{biblio}


\end{document}